\RequirePackage{lineno}
\documentclass[aps,prl,showpacs,twocolumn,groupedaddress]{revtex4}  
\usepackage{graphicx}  
\usepackage{dcolumn}   
\usepackage{bm}        
\usepackage{amssymb}   

\begin{document}
\setpagewiselinenumbers
\modulolinenumbers[5]
\linenumbers

\date{March 26, 2011}
  
\title{ A possible explanation for NuTeV's anomalous dimuon events }
\author{Andrew Alton}
\author{Thomas Alexander}
\affiliation{Augustana College, Sioux Falls, SD, 57197}
\date{\today}

\begin{abstract}
We consider a model where neutrinos interact with weakly interacting massive particles (WIMPs) as a possible explanation for the production of NuTeV's anomalous dimuon events.  This model naturally produces the features reported by NuTeV, specifically the muon energy asymmetry and other kinematics.  When combined with the NuTeV event selection criteria, this model predicts an expectation of zero events as observed in their other search channels. For the best agreement with data, the model requires a lower WIMP mass than conventionally expected, but one that is of current interest due to DAMA/LIBRA results, at around 10 GeV$/c^2$. This model has a very high cross section for neutrino-WIMP interactions.  We make suggestions for future searches that would confirm or refute this model.
\end{abstract}

\maketitle

In the NuTeV experiment, a high energy ($\overline{E}\approx 100 \; {\rm GeV}$) neutrino beam was directed at an instrumented iron target-calorimeter~\cite{Bernstein:1994a,Harris:1999uv}.  The primary goal of the experiment was to precisely measure the Weinberg angle sin$^2\theta_w$~\cite{Zeller:2001hh}.  One of several other experimental goals included searching for neutral heavy leptons (NHLs).  To enhance the sensitivity for NHL searches, a dedicated decay volume (decay channel)  was introduced between the primary target (where neutrinos were produced) and the target-calorimeter (where the neutrinos interact).  

	NuTeV collided 800 GeV protons with a beryllium oxide target.  Secondary particles from the collision were steered by the magnets of the `Sign Selected Quadrupole Train' (SSQT)~\cite{Bernstein:1994a} to a decay region.  The pions and kaons decayed to produce muons and neutrinos.  The sign selection of the SSQT allowed for a very pure beam of $\nu_{\mu}$ (neutrino mode) or $\overline{\nu}_{\mu}$. The pion and kaon decay region was about 350 m long and was followed by a 900 m long shield to absorb everything but the $\nu$'s.  The decay channel and the main detector followed.  
The decay channel consisted of a scintillator veto at the upstream end and drift chambers for tracking.  Between the drift chambers it was filled with vessels of helium to minimize neutrino interactions.  Details of this decay channel and its performance have been published \cite{Vaitaitis:1999wq,Formaggio:1999ne,Adams:2001ska}.  

	One analysis~\cite{Adams:2001ska} of the NuTeV decay channel data searched for final states of $\mu\mu, \; \mu e, $ and $ \mu \pi$. The expected background of all three channels is $0.069 \pm 0.010$ events, however there were three $\mu\mu$ events in the data.  These three events do not match characteristics of NHL's or any standard model sources.  The most viable model proposed to date \cite{Dedes:2001zia,Abazov:2006as} presumes that a neutral massive quasi-stable particle was created at the neutrino target and decayed in the decay channel.  We present an alternative model here.

\begin{figure}
\includegraphics[scale=0.35]{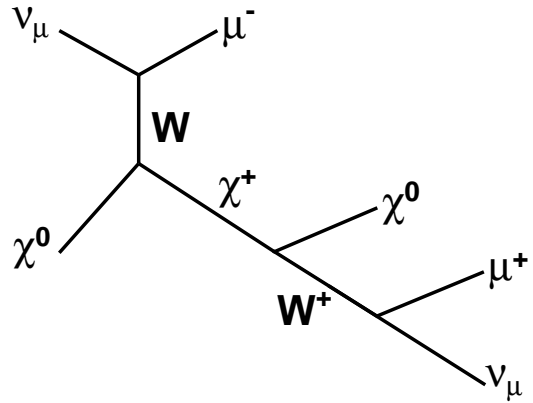}
\caption{
The Feynman diagram for the stand-alone MC.
\label{fig:feyn}}
\end{figure}

Our model is based on the common speculation that a supersymmetric (SUSY) \cite{Haber:1984rc,Nilles:1983ge,Hagelin:1984wv,DeJesus:2004sn} neutralino is a weakly interacting massive particle (WIMP) that accounts for dark matter. Dark matter is hypothesized to explain discrepancies in galactic rotation curves and other astronomical anomalies~\cite{Zwicky:1933gu,Jungman:1995bz,Bennett:2003bz,Komatsu:2008hk}.  Our hypothesis is that the neutrinos in the NuTeV beam interact with neutralinos from the dark matter halo present in the decay channel.  Charged current interactions excite the neutralino ($\chi^0$) to a chargino ($\chi^+$) state and produce a primary muon. When the chargino decays, it can produce a second muon as shown in Fig~\ref{fig:feyn}.  An initial study with Madgraph \cite{Alwall:2007st} revealed that there are hundreds of channels where $\nu_{\mu}+\chi^0\rightarrow\mu^++\mu^-+\nu_{\mu}+\chi^0$.  Madgraph currently assumes the initial particles are massless which is not a valid assumtion for this interaction.

\begin{figure}
\includegraphics[scale=0.4]{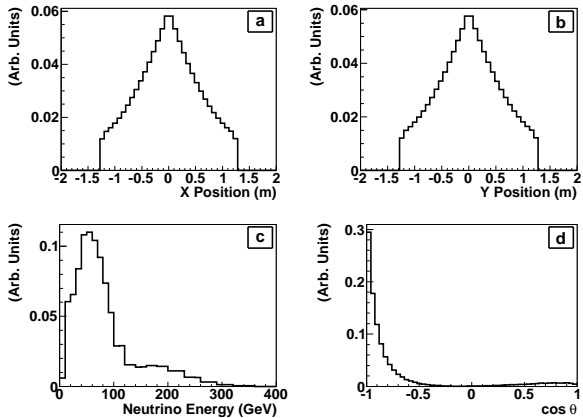}
\caption{
Input distributions for the MC.  The a) x and b) y locations, c) the neutrino energy, and d) the polar angle of the muon in the center of mass frame of the collision.
\label{fig:mcin}}
\end{figure}

We developed a simple stand-alone Monte Carlo (MC) for the process shown in Fig. \ref{fig:feyn}.  Our analysis focuses on neutrino mode running because all three events occurred while NuTeV took data in that mode.  Our model assumes an infinitely short lifetime so as to not spoil the vertexing.  The expected lifetime of $\chi^{\pm}$ is very short and the relativistic effects should be minimal so this assumption seems reasonable. In a given execution of the MC, a specific mass for $\chi^0$ ($M$) and the mass difference between it and $\chi^+$ ($\Delta M$) were used.  The $\chi^0$ was assumed to be at rest.  While this is not generally true, the initial momentum is expected to be negligible.  The interaction location in z was taken to be uniformly distributed, while the neutrino energy and the x and y interaction profile were chosen to agree with the NuTeV larger data set of charged current events.  These inputs are shown in Fig. \ref{fig:mcin}.  The neutrinos are assumed to have no transverse momentum.  In addition, Fig. \ref{fig:mcin}d shows the polar angle of the primary muon in the center of mass frame of the collision.  This angle was extracted from Madgraph as an input to the MC.  

Monte Carlo samples were generated at a variety of values of $M$ and $\Delta M$, and these samples were subjected to the same selection criteria used by the NuTeV analysis.  The selection criteria required exactly two muons with each carrying at least 2.2 GeV of energy, the sum of the energy of the two must exceed 12 GeV (to match the requirement in the other channels that the e or $\pi$ exceed 10 GeV), the angle of each muon was required to be less than 100 mr with respect to the beam direction (ie the z-axis), the muons needed to originate within the fiducial volume of the decay channel, be separated from the location of drift chambers, and each muon must reach the target calorimeter inside a transverse $\pm$1.27 m fiducial area centered on NuTeV's coordinate system.  Finally, there was a transverse mass cut $m_{T}>2.2$ GeV/$c$ where $m_{T}= |P_{T}|+\sqrt{P_{T}^{2}+m_{v}^{2}}$ where $P_{T}$ is the dimuon momentum transverse to the beam direction and $m_{v}$ is the invariant mass of the visible particles.  Reasonable agreement was observed between the NuTeV data and the simulation after event selection criteria were applied.  This agreement will be discussed and shown shortly.

\begin{figure}
\includegraphics[scale=0.4]{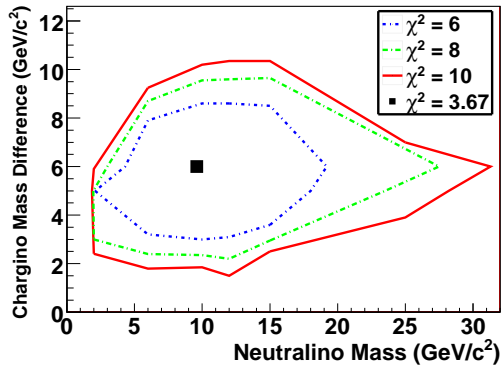}
\caption{
In the region of $\chi^0$ mass ($M$) and $\Delta M$ space with the lowest $\chi^2$, contours of equal $\chi^2$ are shown.  
\label{fig:chi}}
\end{figure}

In order to quantify the agreement between MC samples and data, we defined a $\chi^2$-like metric which used four input distributions: $E_{\mu1}, E_{\mu2},  m_{T} $, and $ z$ of the interaction.  Muon number one is identified by its negative charge when available and the higher energy is used when its charge is not available. For each quantity, the $\chi^2$ compared the mean of the MC events passing selection criteria to the mean of the three NuTeV events and treated the RMS of the NuTeV events as an uncertainty. The $\chi^2$ is defined as
\begin{widetext}
\begin{equation} \chi^2=\frac{( \overline{E_{\mu1data}}-\overline{E_{\mu1MC}})^2}{\sigma_{E\mu1}^{2}}+
\frac{( \overline{E_{\mu2data}}-\overline{E_{\mu2MC}})^2}{\sigma_{E\mu2}^{2}}+
\frac{( \overline{m_{Tdata}}-\overline{m_{TMC}})^2}{\sigma_{m_T}^{2}}+
\frac{( \overline{z_{data}}-\overline{z_{MC}})^2}{\sigma_{z}^{2}}.
\end{equation}
\end{widetext}
A few similar $\chi^{2}$ were tried using different variables.  Specifically, the muon energy asymmetry was included in some definitions of $\chi^2$, and some $\chi^2$'s did not include the quantities $m_{T}$ or z.  The set in Eq. 1 seems to include all variables with strong sensitivity and no varriables that are strongly correlated to others.  By quantifying the agreement in this way, we were able to generate MC samples and scan a large range of values for the masses of $\chi^0$ ($M$) and $\Delta M$. We found that the NuTeV data gave a minimum $\chi^2$ of 3.47 with 4 degrees of freedom at $M$ = 9 GeV/c$^2$ with  $\Delta M$ = 6 GeV/c$^2$as you can see in Fig.~\ref{fig:chi}. The range of scanned $\chi^0$ mass was from 0.5 GeV/c$^2$ to 1500 GeV/c$^2$; while, $\Delta M$ was scanned from 1 GeV/c$^2$ to 30 GeV/c$^2$.  
 
\begin{figure} 
\includegraphics[scale=0.4]{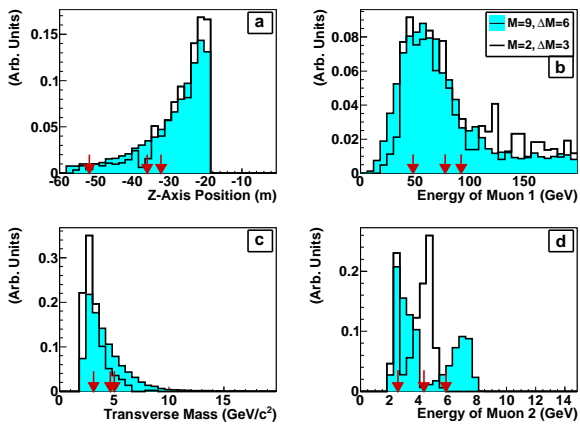}
\caption{
The input distributions for the $\chi^2$ at its minimum ($M$ = 9 GeV/c$^2$ with a $\Delta M$ = 6 GeV/c$^2$) and at a lower $\Delta M$ value ($M$=2 GeV/c$^2$ with a $\Delta M$ = 3 GeV/c$^2$).  The arrows show the values of the 3 NuTeV events.  The four distributions are a) z position, b) $E_{\mu1}$ of the negative muon, b) $m_{T}$  and d) $E_{\mu2}$.
\label{fig:inputs}}
\end{figure}

\begin{figure}
\includegraphics[scale=0.4]{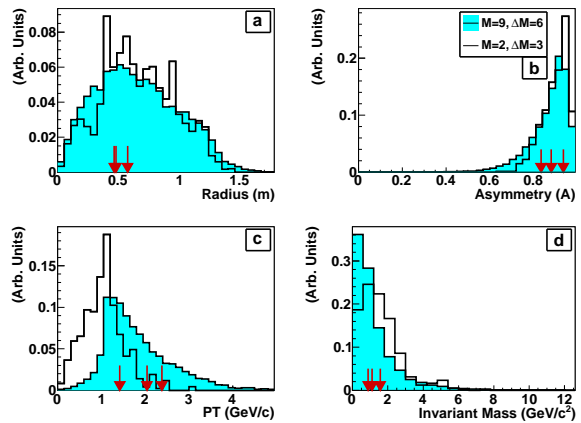}
\caption{
Four other kinematic distributions where the $\chi^2$ is at its minimum (M$_{\chi^0}$=9 GeV/c$^2$ with a $\Delta M$ of 6 GeV/c$^2$) and at a lower $\Delta M$ value ($M$ = 2 GeV/c$^2$ with a $\Delta M$ = 3 GeV/c$^2$).  The arrows show the values of the three NuTeV events.  The four distributions are a) the transverse radius of the interaction, b) the muon energy asymmetry,  c) the total $p_{T}$ of the dimuon pair and d) invariant mass of the dimuons.
\label{fig:kinem}}
\end{figure}

The NuTeV events show reasonable agreement with the variables used in the $\chi^2$ as shown in Fig. ~\ref{fig:inputs}.  The $E_{\mu2}$ shape is due to two categories of events being accepted by the selection criteria.  In the rest frame of the W$^+$, only a $\mu^+$ going approximately parallel or anti-parallel to the rest/lab boost direction will be accepted by the selection criteria.  Those muons with a substantial transverse component have a negligible chance of being accepted.  
It is possible that this structure may allow additional sensitivity compared to using just the mean; however, we did not explore this possibility.  The NuTeV events also show agreement with other kinematic variables in Fig ~\ref{fig:kinem}.  Specifically, NuTeV events sit firmly in the bulk of the distribution for the transverse radius of the interaction, the muon asymmetry (A=$\frac{E_{\mu1}-E_{\mu2}}{E_{\mu1}+E_{\mu2}}$),  the total $p_{T}$ of the dimuon pair and invariant mass of the dimuons.
   
The acceptance of the selection criteria applied to this model is very low.  At the $\chi^2$ minimum, the acceptance is $2.1 \times 10^{-4}$, growing by about an order of magnitude as the mass of the $\chi^0$ decreases to 2 GeV/$c^2$ and getting smaller as the mass increases.  
This behavior is qualitatively expected: the larger the mass, the larger the possible opening angle between the muons, and the less likely both muons will hit NuTeV's main detector.  This effect is also responsible for the steep shape of z for the accepted MC events, as shown in Fig ~\ref{fig:inputs}a.   The low acceptance also means that, if this is an accurate model, other signatures should be observable.  These will be discussed later. 

One feature of this analysis, which is unexpected in WIMP interactions, is the cross section.  One would expect a cross section within a few orders of magnitude of $10^{-45}$ cm$^{2}$. However, when one considers the relatively small ($10^{14}$) number of neutrinos, the number of WIMP targets ($10^{6}$) and the poor acceptance, in order to generate the three events in NuTeV, we would arrive at a cross section of around $10^{-20}$ cm$^{2}$.  While this is one of the least agreeable features of our model, it does make independent confirmation or rejection very feasible. 

In addition to the cross section, the fact that low-mass charginos and neutralinos were not observed by the LEP experiments in their examination of the Z width~\cite{Drees:2001xw} raises questions with our model. Some authors (for example~\cite{Kuflik:2010ah}) have found regions of SUSY space that may still allow a neutralino with a mass like ours.  To this point, our analysis has not made use of any SUSY specific characteristics.  So, to minimize these concerns, we will generalize our model.   Our charged state is not necessarily a chargino, and our neutral state is not necessarily a neutralino, and, possibly, the mediator of the charged current reaction is not the standard model W. To minimize confusion throughout the rest of this paper, we will call the neutral state C$^0$, the charged state C$^{\pm}$, and the mediator D$^{\pm}$. The C$^0$ and C$^{\pm}$ cannot interact with the Z boson with standard couplings.  

Dark matter candidates in the mass range that NuTeV's data prefer in our model have received a lot of attention lately
(for example~\cite{Petriello:2008jj,Draper:2010ew,Feldstein:2010su,Barger:2010mc, Belanger:2010cd,Hooper:2010uy} and many of their citations) in response to the DAMA/LIBRA~\cite{Bernabei:2008yi} results. While they have interpreted the strong annual modulation of their events as dark matter interactions, other experiments~\cite{Ahmed:2009zw} have excluded their results.   A feature of crystals, called channeling~\cite{Feldstein:2009np}, has led some~\cite{Bottino:2009km,Aalseth:2010vx} to reevaluate the results, finding a small mass region between ~7-11 GeV/$c^2$ or  ~3-8 GeV/$c^2$~\cite{Petriello:2008jj} for the WIMP.

A final concern is charged particle production (C$^+$C$^-$) at LEP.  Actual chargino production at LEP may occur through s-channel production via the Z or $\gamma$ and through t-channel sneutrino exchange.  However, these modes destructively interfere\cite{Bartl:1991re}.  In our model, the C$^{\pm}$ is not allowed to couple to the Z. It is reasonable to suggest that our C$^0$, or another as yet unhypothesized particle, could participate in a t-channel interaction and the destructive interference with s-channel $\gamma$ production may be more complete, producing a low enough cross section ($\sim$0.3 pb) to escape detection.  Pair-production of our C$^{\pm}$ may also have escaped detection at LEP because recent searches (for example \cite{Alexander:1996nf,Abbiendi:2003sc}) for charginos are only sensitive above a $\Delta M$ of at least 3.0 GeV/c$^2$.  A reasonable portion of our favored space is available with $\Delta M$ below 3.0 GeV/c$^2$.   For the simulated sample at $M$ = 9 GeV/c$^2$ and $\Delta M$ = 3 GeV/c$^2$, the distribution of $E_{\mu2}$ values are always below 3.5 GeV, in stark disagreement with two of the NuTeV events.  The simulation at $M$ = 2 GeV/c$^2$ and $\Delta M$ = 3 GeV/c$^2$ has a good $\chi^2$ of 7.7 and a reasonable visual agreement.  The distributions from this sample are shown in Figs.~\ref{fig:inputs} and~\ref{fig:kinem}. 

Because multiple options may have allowed our model to have gone unnoticed until now, and, given the current interest in dark matter of order 5 GeV/c$^2$, we have examined our model for ways in which the NuTeV collaboration may be able to confirm or refute this hypothesis. Instead of simulating backgrounds, we suggest event topologies for which the acceptance was relatively large and where we expect the backgrounds to be reasonable.  One might think that NuTeV's examination \cite{Adams:1999mn} of low hadronic energy dimuons in their main detector should have seen events caused by neutrino-neutralino scattering, however a few factors compete with better detector coverage.  In that analysis, the lower energy muon was required to have an energy greater than 5 GeV and a positive z momentum (P$_z$).  In addition, NuTeV's main detector has a smaller fiducial volume.  These lead us to expect fewer than 10 events due to our proposed model. If the energy requirement on this channel were lowered to 2.2 GeV, about 3 times as many signal events would be expected.

There are a number of ways NuTeV could loosen their requirements and see more events in the decay channel of their detector.  One option would be to require two tracks to be reconstructed, but allow one track to miss the detector.  The track that misses the detector would still be required to hit two drift chambers and reconstruct a good vertex.  With these requirements, our model would expect between 20 (at low mass) and 45 (at high mass) events.  It turns out NuTeV did not use the region less than 19 cm from the face of the main detector for the previous analysis.  Presumably, this was in order to have multiple drift chamber hits from which to form tracks before the main detector (primarily for electrons and pions).  A second option, since we are interested in muons that can be found in the main detector, is to search, only the region from 19 cm to 2 cm from the face.  For this search our model expects about 10 events with both muons reconstructed in the main detector.  If we also remove the requirements that both muons have an angle of less than 1 mrad we would expect around 25 events.   A third option would be to remove the requirements on $m_{T}$, the angle and the energy of the second track, but still require both tracks to reach the main detector.  NuTeV almost certainly required the muons energy above 2.2 GeV to reduce standard model backgrounds when they were looking for a couple of events.  However, with this selection we would expect at least 17 neutrino-neutralino scattering events.  

  A clear signature of these events would be a pair of muons where one is a backward going muon in the lab frame. Backward going ($P_z<0$) muons in the decay channel need to be restricted to regions where each muon hits at least two drift chambers, which produces a much smaller fiducial length and the acceptance is not much larger than their previous result. However, in the main detector, backwards going muons are an interesting option. The backward going muons have very low energy, so good acceptance requires lowering the energy threshold to 1.1 GeV, which is still enough for significant penetration. If we look in the main detector for one muon with an energy greater than 9 GeV moving in the forward direction, one muon with an energy greater than 1.1 GeV and a z component of momentum less than -0.02 GeV/$c$, and a dimuon energy greater than 12 GeV, then we only expect one neutrino-neutralino scattering event at $M$=2 GeV/$c^2$.  But, by $M = 6$ GeV/$c^2$ the expectation has risen to 140 events and it stays at this level or above all the way to $M=75 $ GeV/$c^2$. If backgrounds require stronger constraints, one could insist that the backward muon penetrate at least four drift chambers (0.8 m of iron) and the forward muon have it's momentum and charge analyzed in the toroid.  Then, we expect no events at very low mass ($M = 2$ GeV/$c^2$) but for $M =10 $ GeV/$c^2$ and above we expect 80 neutrino-neutralino scattering events.

  In addition to NuTeV, a number of past, current, and future experiments should be able to search for similar topologies. At our lowest  $\chi^2$, the neutrino energy threshold is 10 GeV, so MINOS or MINER$\nu$A would likely only be able to study this if they ran the NUMI beam in high energy mode.  CHORUS and NOMAD both have high enough energy, should have enough intensity, and have looked at dimuon production before \cite{Kayis:2008xp,Astier:2000us}.  NOMAD has studied backward going particles \cite{Astier:2001wd}.  Given these experiences, they should be able to look for clean two muon events with one muon going backwards and perhaps quickly rule out this explanation for the NuTeV experiment's dimuon excess.  In the future, if the NuSoNG \cite{Adams:2009gf} experiment were brought to life, they would expect 60 events with the same topology.  By instrumenting the sides of their helium region, they ought to be able to quickly confirm or refute this explanation.  
  
    One might also ask why the current collider detectors have not noticed signs of new physics due to this model.  It is possible they already have. In Refs. \cite{Aaltonen:2008qz,Aaltonen:2010zz}, the CDF collaboration reports observing a significant number of relatively low momentum muons which appear to originate very far from the interaction point. If our C$^{\pm}$ has a lifetime somewhat larger than a B meson's, then pair production of C$^{\pm}$ would appear this way.  Alternatively, b and c quarks produced in interactions, would decay producing neutrinos which would interact with dark matter in the detector volume.  However, the D0 collaboration searched and was unable to confirm this signal\cite{Williams:2009uq}. In addition to pair production of the charged state, hadron colliders could produce a mixed state with C$^{\pm}$C$^0$.  That assumes, however that the D$^{\pm}$ interacts with quarks. Pair production of the neutral state assumes some Z-like object which interacts with the neutral state and quarks. If these final states can be produced, we would need a value for the coupling to quarks in order to compare to the Z + jet and $\mu$ + MET topologies that will conceal them. 
    
  Our hypothesis has a number of attractive features in that it provides an explanation for the NuTeV dimuon excess, and the kinematics, especially the asymmetry, are in reasonable agreement with NuTeV's events.  Our hypothesis is also in agreement with the feature that all of NuTeV's events were in neutrino mode, because the integrated neutrino beam had about 3 times as much intensity as the anti-neutrino.  It is a natural result of the selection criteria in this model that there will be no $\mu e$ and no $\mu \pi$ events because the flavor of the high energy lepton is determined by the incident neutrino flavor, and the particle due to the decay is always below the $e$ or $\pi$ threshold of the NuTeV analysis.  In addition, this hypothesis has the interesting feature that it indicates a low mass dark matter candidate.
   
  We have considered a simple model to explain NuTeV's anomalous dimuon events.  In this model, the neutrinos undergo charged current scattering off dark matter particles and excite them to a more massive charged state.  When the massive charged state decays back into the dark matter particle, the net visible signature is two muons.  This model nicely describes the kinematics and other features of the NuTeV events, and those kinematics are used to extract a value for the mass of the dark matter particle and the mass of its charged partner. The most unsettling features of this model are the cross section, which is many orders of magnitude above expectations, and the unobserved low mass charged particle, C$^{\pm}$.  We have provided a number of suggestions for further studies which could confirm or reject this explanation.    
  
  We thank the NuTeV experiment and Fermilab staff for their original work.  
We thank Todd Adams, Don Lincoln, Tim Bolton, Bob Bernstein, Janet Conrad, Eric Wells, and Nathan Grau for useful comments and discussion.
  The authors were supported by the South Dakota Space Grant Consortium, Research Corporation, the Augustana Research and Artists Fund, and the State of South Dakota CUBED center.

\end{document}